\documentstyle[aps, epsf]{revtex}

\newcommand{\pa}{\partial}

\begin{document}

\draft
\title{\hspace{14cm}{\rm {\normalsize{USTC-ICTS-03-06}}}\\
\vspace{1cm} REMARKS ON 't HOOFT'S BRICK WALL MODEL}
\author{Hua Bai\footnote{E-mail address: huabai@mail.ustc.edu.cn}}
\address{Interdisciplinary Center for Theoretical Study,
University of Science and Technology of China\\
Hefei, Anhui 230026, P.R. China}
\author{Mu-Lin Yan\footnote{E-mail address: mlyan@staff.ustc.edu.cn}}
\address{CCAST(World Lab), P.O. Box 8730, Beijing, 100080, P.R. China \\
  and \\
Interdisciplinary Center for Theoretical Study,
University of Science and Technology of China\\
Hefei, Anhui 230026, P.R. China\footnote{mail address}}
\date{\today}
\maketitle

\begin{abstract}
{ A semi-classical reasoning leads to the non-commutativity of the
space and time coordinates near the horizon of Schwarzschild black
hole. This non-commutativity in turn provides a mechanism to
interpret the brick wall thickness hypothesis in 't Hooft's brick
wall model as well as the boundary condition imposed for the field
considered. For concreteness, we consider a noncommutative scalar
field model near the horizon and derive the effective metric via
the equation of motion of noncommutative scalar field. This metric
displays a new horizon in addition to the original one associated
with the Schwarzschild black hole. The infinite red-shifting of
the scalar field on the new horizon determines the range of the
noncommutativ space and explains the relevant boundary condition
for the field. This range enables us to calculate the entropy of
black hole as proportional to the area of its original horizon
along the same line as in 't Hooft's model , and the thickness of
the brick wall is found to be proportional to the thermal average
of the noncommutative space-time range. The Hawking temperature
has been derived in this formalism. The study here represents an
attempt to reveal some physics beyond the brick wall model. }
\end{abstract}
\vspace{0.3in}

The brick wall model proposed by 't Hooft has been used for the
purpose of deriving  the entropy of black hole and other
quantities \cite{thooft85}\cite{thooft96}, and has been
extensively studied (an incomplete list, see
Refs.\cite{exp}\cite{mlyan}). In the model, the thickness of the
brick wall near the horizon of Schwarzschild black hole was set to
be
\begin{equation}
 h={N'l_p^2\over 360\pi r_H}
\end{equation}
where $r_H$ is the radius at which the horizon is located,
$l_p=\sqrt G$ (in this letter $\hbar=c=1$) the Planck length and
$N'$ the number of the multiplet of the quantum field in the
model. Eq.(1) is {\it a prior} hypothesis of the brick wall model.
Actually, only under this hypothesis, the thermodynamic properties
of a black hole can be reproduced correctly. Namely,  this model
can lead to the correct Bekenstein-Hawking entropy formula
\begin{eqnarray}\label{ee}
S_{BH}={1\over 4}{A\over G}
\end{eqnarray}
where $S_{BH}$ is Bekenstein-Hawking entropy
\cite{Bekenstein72}\cite{Hawking74} and $A$ is the horizon area .
In this letter, we try to derive the brick wall thickness by a
semi-classical argument  and to reveal some underlying physics
related to this hypothesis.

For the sake of definiteness, we study the $3+1$ dimensional
Schwarzschild black hole. In this case, $\pa_t$ is the time
Killing vector. Its global energy (or the mass of the black hole)
is $E_{BH}=M=r_H/2G$. Rather than thinking black hole as a
classical object, we treat it as a quantum state with high
degeneracy and its degrees of freedom are located near by the
horizon. Treating the energy $E_{BH}$ and its conjugate time
coordinate as operators, quantum mechanics tells that these two
quantities  can not be measured simultaneously. In other words,
 $E_{BH}$ and t satisfy Heisenberg uncertainty relation $[t,E]= i$.
Given the relation $E_{BH}=r_H/2G$, we must conclude that the
uncertainty of $E_{BH}$ imply that of $r_H$, and then we have
\begin{eqnarray}
[t,r]|_{r=r_H}=i{2 l_p^2 },
\end{eqnarray}
  This equation implies that the radial
coordinate is noncommutative with the time at the horizon. The
corresponding uncertainty relation for them is $(\Delta t)(\Delta
r)|_{r\sim r_H}\sim 2l_p^2$. In other words, due to quantum
measurement effects, $r$ spreads in the range of $(r-\Delta r,
r_H+\Delta r)$.  Namely, eq.(3) is extended as follows
\begin{eqnarray}\label{e4}
[t,r]\mid_{r\in (r_H-\Delta, r_H+ \Delta)}=i{2 l_p^2 }.
\end{eqnarray}
where $\Delta$ is a distance about Planck length scale. Here, for
simplicity, we have denoted on
 $\Delta r = \Delta$ with $\Delta$ on the order of $l_p$. For a classical observer at infinity,
 he can only detect events happening at $r > r_H$. For this reason, we only need to consider
$r > r_H$  at least semi-classically. In the range of $(r_H,
r_H+\Delta )$, the classical $\phi^a-$fields are noncommutative. The
non-commutative theory in field theory and in string theory has
been discussed recently in \cite{Witten,Seiberg}. As
$r>r_H+\Delta$, the fields are commutative as usual since
\begin{equation}\label{00}
[t,r]\mid_{r>r_H+\Delta}=0.
\end{equation}
According to our aforementioned discussion, we should construct a
brick wall model with noncommutative $\phi^a-$fields in the range of
$(r_H, r_H+\Delta )$ and with commutative $\phi^a-$fields in
$(r_H+\Delta, L)$ where $L$ represents an infrared cutoff in the
model. It is essential that $\Delta$ should be an intrinsic
quantity of the model, which characterizes the boundary between the
noncommutative space-time range and the commutative space-time range,
and should be determined by the
model itself consistently. Surprisingly, this
expectation can be realized  and a brick wall
model without the brick wall thickness hypothesis can be constructed
by the following considerations: 1)
Starting with a simplest noncommutative $\phi^a-$field action
within metric of Schwarzschild black hole, the equation of motion
of $\phi^a$ can be derived exactly; 2) This equation of motion in
noncommutative field theory should be, of course, quite different
from the ordinary Klein-Gordon equation of $\phi^a$-fields within
Schwarzschild metric. This fact implies $\phi^a-$fields should be moved in a
curve space with a new effective metric $\widetilde{g}^{\mu\nu}$
for a classical observer at infinity;
3) Remarkably, it will be shown below that $\widetilde{g}^{tt}$ has
two new singularities besides the original one at $r=r_H$ : one is outside the horizon of the black
hole and another is in inside. Denoting their locations as $r_H
\pm \Delta'$ respectively, we will find $\Delta'$ is dependent on
the energies of the noncommutative fields $E$, i.e.,
$\Delta'=\Delta'(E)$. This means that to the fields with energy
$E$, $\phi^a_E(r)$, its red-shifting on the $(r_H \pm
\Delta'(E))$-surface  is infinite due to
$\widetilde{g}_{00}(r=r_H+\Delta(E))=0$, and then we have
\begin{eqnarray}\label{000}
\phi^a_E(r)|_{r=r_H+\Delta'(E)}=0;
\end{eqnarray}
4) We argue that the fact that the noncommutative fields vanish at
$r=r_H+\Delta'(E)$ means we can think the space-time coordinates
on the surface of $r=r_H+\Delta'(E)$ to be commutative , i.e.,
$[t, r]|_{r=r_H+\Delta'(E)}=0$. And then we will further have
\begin{equation}\label{01}
[t, r]|_{r\geq r_H+\Delta'(E)}=0.
\end{equation}
Comparing eq.~(\ref{01}) with ~(\ref{00}), we get
\begin{equation}\label{02}
\Delta'(E)=\Delta.
\end{equation}
Namely, $\Delta=\Delta(E)$ can be determined by using
eqs.~(\ref{000}) and ~(\ref{02}), and the ordinary brick wall model
works in the raging of $(r_H+\Delta, L)$, whose ultraviolet boundary
condition is eq.~(\ref{000}).
Consequently, we have constructed a new model
without the brick wall thickness hypothesis by starting with thinking
the black hole as a quantum state with high degeneracy. In the follows, we shall
apply this model to derive the entropy of the Schwarzschild black hole, and to
interpret 't Hooft's hypotheses on the brick wall thickness.

With the above, we now consider the classical field as a probe in
the region of $r_H < r < r_H + \Delta(E)$ but moving in a
non-commutative background. We rewrite eq.~(\ref{e4}) as follows
\begin{eqnarray}\label{e9}
&&[x^i,x^j]=i\Theta\varepsilon^{ij},\;\;\;
(i,j=0,1),\;\;\;(x^0=t,x^1=r),\;\;\;
 \Theta={2l_p^2}, \\
 \label{e10}
&&[x^k,x^\mu]=0, \    \ (k=2,3;\mu=0,1,2,3)
\end{eqnarray}
 where $\varepsilon^{ij}$ is an antisymmetrical tensor with $\varepsilon^{01} = 1$.
 The star product of two function $f(x)$
and $g(x)$ is given by the Moyal formula:
\begin{equation}\label{sp}
(f\star g)(x)= \exp\left[{i\over
2}\Theta\varepsilon^{ij}\frac{\partial}{\partial
x^i}\frac{\partial}{\partial y^j}\right]f(x)g(y)|_{y = x}.
\end{equation}

 For simplicity, we consider only the coupling between the
 field and the background, i.e.,
\begin{equation}\label{e12}
 {I=-{1\over 2}\int{d^{4}x{\sqrt{-g}}g^{\mu\nu}(\partial_\mu
\phi_E\star
\partial_\nu\phi_E)}} \;\;\;\;\;\;\;\;\;\;\;\;(r_H<r<r_H+\Delta)
\end{equation}
where $g_{\mu\nu}$ is the metric of a Schwarzschild black hole,
\begin{equation}
{{ds^{2}}=-\left(1-{r_H\over r}\right)dt^{2}+\left(1-{r_H\over
r}\right)^{-1}dr^2+r^{2}(d\theta^2+\sin\theta^2d\varphi^2)}.
\end{equation}

We evaluate the start product in the action using Eq. (\ref{sp})
and cast the action in the ordinary product. By this,  the
noncommutative effect can be absorbed into an equivalent
background metric. In other words, we first take the
semi-classical quantum effect into consideration. This effect is
then realized through the non-commutative geometry. Finally, this
effect is further through an effective background but in an
ordinary geometry.  For a given energy mode, i.e., assuming
$\phi_E (t,r,\theta,\varphi)=e^{-itE}f (r,\theta,\varphi)$, the
effective metric can be either read from the action (actually
simpler) or from the following equation of motion for the scalar
field once the star product is evaluated:
\begin{eqnarray}\label{e15}
&&\pa_t(\sqrt {-g}g^{tt}\pa_t\phi^a)+\pa_r(\sqrt {-g}
g^{rr}\pa_r\phi^a)+\pa_\theta(\sqrt
{-g}g^{\theta\theta}\pa_\theta\phi^a)+\pa_\varphi(\sqrt
{-g}g^{\varphi\varphi}\pa_\varphi\phi^a) + \nonumber \\
&&{1\over 2!}\left({i\Theta\over 2}\right)^2(\sqrt
{-g}g^{rr}),_{rr}\pa_t\pa_t\pa_r\pa_r\phi^a+\sum_{n=1}^\infty{1\over
2n!}\left({i\Theta\over 2}\right)^{2n}(\sqrt
{-g}g^{tt}),_{\underbrace{r\cdots
r}\limits_{2n}}\underbrace{{\pa_t\cdots
\pa_t}}\limits_{2n+2}\phi^a=0
\end{eqnarray}
where $(\sqrt{ -g}g^{tt}),_{\underbrace{r\cdots r}\limits_{n}}$
stands for $\underbrace{{\pa_r\cdots \pa_r}}\limits_{n}(\sqrt{
-g}g^{tt})$, etc. For the scalar field with given energy $E$, the
above equation becomes
\begin{eqnarray}\label{e16}
&&\left[-{\sin\theta r^3\over r-r_H}-{\sin\theta\Theta^2 E^2\over
4}-{\sin\theta r_H^3\over r-r_H}\sum_{n=1}^\infty
\left({\Theta E\over 2(r-r_H)}\right)^{2n}\right]\phi_E^a,_{tt}\\
\nonumber &&+\left[\sin\theta r(r-r_H) + {\sin\theta\Theta^2
E^2\over
4}\right]\phi_E^a,_{rr}+\pa_\theta(\sin\theta\pa_\theta\phi^a)+{1\over
\sin\theta}\phi_E^a,_{\varphi\varphi}=0
\end{eqnarray}
Noticing
\begin{eqnarray}
\sum_{n=1}^\infty ({\Theta E\over 2(r-r_H)})^{2n}={1\over
1-{\Theta^2 E^2\over 4(r-r_H)^2}}-1,
\end{eqnarray}
we can read the effective metric $\tilde g_{\mu\nu}$ from Eq.
(\ref{e16}) as
\begin{eqnarray}
&&-{\sin\theta r^3\over r-r_H}\left[1+{\Theta^2 E^2(r-r_H)\over
4r^3}+{r_H^3\over r^3}\left({1\over 1-{\Theta^2 E^2\over
4(r-r_H)^2}}-1\right)\right]=\sqrt{-\widetilde{g}}{\widetilde{g}^{tt}}, \\
&& \sin\theta[\left[r(r-r_H)+ {\Theta^2 E^2\over
4}\right]=\sqrt{-\widetilde{g}}{\widetilde{g}^{rr}}, \\
&& \sin\theta=\sqrt{-\widetilde{g}}{\widetilde{g}^{\theta\theta}}, \\
&& {1\over
\sin\theta}=\sqrt{-\widetilde{g}}{\widetilde{g}^{\varphi\varphi}}.
\end{eqnarray}

 From the above, we can solve the effective metric as
\begin{eqnarray}
 &&\sin\theta^2 {\widetilde{g}_{\theta\theta}}=
 {\widetilde{g}_{\varphi\varphi}},\\
 &&{\widetilde{g}_{tt}}{\widetilde{g}_{rr}}=-1 ,\\
 &&{\widetilde{g}_{tt}}=-(1-{r_H\over r})\sqrt{{1+{\Delta^2(E)\over
 r(r-r_H)}}\over{1+{\Delta^2(E)(r-r_H)\over r^3}+{r_H^3\over
 r^3}[{1\over 1-{\Delta^2(E)\over (r-r_H)^2}}-1]}},\\
 &&\tilde g^2_{\theta\theta} = \left[r (r - r_H) + \Delta^2
 (E)\right] \left[r^2 + r r_H + r_H^2 + \Delta^2(E) + \frac{(r -
 r_H) r_H^3}{(r - r_H)^2 - \Delta^2 (E)}\right],
 \end{eqnarray}
where we have set $\Delta (E) = \Theta E/2$. This effective metric
is quite different from the original one. The semi-classical
effect causes the unexpected appearance of a new horizon at $r =
r_H + \Delta (E)$ at which $\tilde g_{tt}$ vanishes. Note also
that the $\tilde g_{\theta\theta}$ blows up at this point which
implies that the curvature scalar vanishes at this point, too.
therefore a regular one. Note that, as also discussed in
\cite{thooft96}, the energy $E$ for the scalar $\phi$ shouldn't be
too large, therefore $\Delta (E) = \Theta E/2$ is not larger than
the Planck length.

With the above, it is now not difficult to understand the boundary
condition (\ref{000}) (with (\ref{02})) we given earlier. Because of the appearance
of the new horizon, an observer cannot detect the $\phi_E$ at $r
\le r_H + \Delta (E)$. Just we have discussed earlier that the field has an infinite
red-shift at $r = r_H + \Delta(E)$. And since the noncommutative
property of $\phi_E^a$ is caused by $[t,r]\neq 0$,
$\phi_E^a(r_H+\Delta(E))=0$ means $[t,r]=0$ at $r=r_H+\Delta(E)$,
as $r > r_H + \Delta (E)$, the field $\phi_E$ should be a
commutative one as discussed earlier. This in turn provides an
explanation for the short-distance cutoff, or the brick wall
thickness $h$, introduced in 't Hooft's model. In other words, the $h$
has its origin of our $\Delta(E)$. We emphasize again that there is no
brick wall in our model, and the $\Delta(E)$ is derived by the action (\ref{e12}).
 With this understanding, we can
now follow 't Hooft's method \cite{thooft96} to evaluate the black
hole entropy in our model. For this purpose, we consider $N$
scalar fields each with energy $E$ in the original Schwarzschild
black hole background. The action in the range of
$(L>r>r_H+\Delta(E))$ reads
\begin{eqnarray}
I={-{1\over 2}\int{d^{4}x{\sqrt{-g}}g^{\mu\nu}(\partial_\mu
\phi^a_{E}\partial_\nu\phi^a_{E})}},\    \     \     \
(r>r_H+\Delta(E)).
 \end{eqnarray}
The eq.~(\ref{000}) serves as the ultraviolet boundary condition
of $\phi_E^a$. And according to refs\cite{thooft85,thooft96}
another boundary condition can be taken as $\phi_E^a(r=L)=0$,
where $L$ is large.

 Using the $WKB$ approximation and by setting
$\phi_E^a=e^{-iEt-i\int k(r)dr}Y_{lm}(\theta,\varphi)$, we have
the radial wave number $k(r,l,E)$ from the corresponding equation
of motion:
 \begin{eqnarray}
 \label{nm}
 k^2=\left(1-{r_H\over r}\right)^{-1}\left[\left(1-{r_H\over r}\right)^{-1} E^2  -
 r^{-2}l(l+1)\right], \;\;\;{\rm in\;the\;range\;of\;}(
 L>r>r_H+\Delta(E)).
 \end{eqnarray}
 The number of states below energy $E$ is
 \begin{eqnarray}
 g(E) &=&N\int dl(2l+1)\int_{r_H+\Delta(E)}^L
 dr\sqrt{k^2(l,E)} \nonumber \\
 &=&{2N\over 3}\int_{r_H+\Delta(E)}^L dr {r^4 E^3\over
 (r-r_H)^2}.
 \end{eqnarray}
 Note that be different from 't Hooft's brick wall model, there is no
 an additional ultraviolet cutoff to be introduced in above calculation of $g(E)$.
 The free energy then reads
 \begin{eqnarray}
 \pi\beta F &=& \int d g(E) \ln(1-e^{-\beta E})\nonumber \\
 &=&-\int_0^\infty dE {\beta g(E)\over e^{\beta E}-1}\nonumber \\
 &=&-{2N\over3}\int_0^\infty d E {\beta E^3\over e^{\beta E}-1} \int_{r_H+\Delta(E)}^L
 dr{r^4\over (r-r_H)^2}.
 \end{eqnarray}
 The dominant contribution from the event horizon to $F$ is
 \begin{eqnarray}
 F\approx -{8N\zeta(3) r_H^4\over 3\pi\Theta \beta^3}
 \end{eqnarray}
 where we have used $\Delta(E) = \Theta E/2$  and $\zeta(3)$ is Riemann
 $\zeta$-function, $\zeta(3)= {\sum_{n=1}^\infty}1/n^3 \approx 1.202.$

 The entropy of the black hole can now be obtained as
  \begin{eqnarray}
  \label{S}
  S = \beta^2{\pa F\over \pa \beta}={8N\zeta(3)r_H^4\over
  \pi\Theta\beta^2}.
 \end{eqnarray}

Now, let us derive the inverse temperature $\beta$ in the above
equation in the formalism of this present paper. There are great
deal of works in the literature to study and to discuss the black
hole's temperature $\beta^{-1}$ (e.g., see
\cite{Hawking74}\cite{Ruffini}-\cite{Wilczek}). We employ the
methods of Damour-Ruffini\cite{Ruffini} and of Sannan\cite{Sannan}
to do so. In this method, the point is that we should derive the
outgoing wave functions of $\phi_E$ in both outside and inside of
the black hole. Using the $\phi_E$-wave-number expression
(\ref{nm}), the incoming wave function of $\phi_E$
 reads
\begin{eqnarray}
\phi^{in}_E=\exp\left[{{-iEt-i\int
k(r)dr}}\right]=\exp\left[{{-iE(t+\int^r_c {r\over
(r-r_H)}dr)}}\right].
\end{eqnarray}
In the above expression, the integration $\int^r_c {r/
(r-r_H)}dr=r-c+r_H \ln ((r-r_H)/(c-r_H))$ is actually of the so
called tortoise coordinate $r^*=r+r_H\ln ((r-r_H) /r_H)$ used in
ref.\cite{Ruffini}, expect an unimportant constant. For
simplicity,  we take $c=r_H+\Delta(E)$, and set
$t+\int^r_{r_H+\Delta(E)} r/(r-r_H)dr =v$ which is the usual
advanced Eddingtion-Finkelstein coordinates. Then the incoming
wave function is $\phi^{in}_E=e^{-iEv}$, and the outgoing wave
function in the range $r>r_H+\Delta(E)$ is
\begin{eqnarray}\label{out1}
\phi^{out}_E(r>r_H+\Delta(E))=A_E\exp
\left[{-iEt+i\int^r_{r_H+\Delta(E)} {Er\over (r-r_H)}dr}\right]
=A_E\phi^{in}_E\exp \left[{2i\int^r_{r_H+\Delta(E)} {Er\over
(r-r_H)}}dr\right]
\end{eqnarray}
where $A_E$ is normalized constant. From the equation(\ref{e16})
which holds in the range of $r_H-\Delta(E)<r<r_H+\Delta(E)$ (see
eq.(\ref{e4})), the square of $\phi_E$-wave- number $k'^2(r)$
reads
\begin{eqnarray}
k'(r)^2={E^2r^2\over (r-r_H)^2+{(r-r_H)\Delta^2(E)\over
r}}\left[1+{(r-r_H)\Delta^2(E)\over r^3} +{r_H^3\over
r^3}\left({(r-r_H)^2\over
(r-r_H)^2-\Delta^2(E)}-1\right)-{l(l+1)(r-r_H)\over
E^2r^3}\right].
\end{eqnarray}
By using S-wave approximate, the wave number $k'$ reduces to be
\begin{eqnarray}\label{wave}
k'(r)=\pm i {Er_H\over
\Delta(E)\sqrt{1-x^2}}\sqrt{x+3\xi(x^2-1)+3\xi^2x(x^2-1)+\xi^3(x^4-1)\over
{x(1+x\xi)(1+{\xi\over \xi x^2+x})}}
\end{eqnarray}
where $x=(r-r_H)/\Delta(E)$ and $\xi=\Delta(E)/r_H$. Note, the
mass of the black hole we concern is much larger than the Planck
mass, hence $\xi\ll 1$. Considering the continuity of wave
functions of $\phi_E^{out}$, and using eq.(\ref{out1}), the
outgoing wave function in the inside of the black hole,
$\phi_E^{out}(r<r_H-\Delta(E))$, is obtained as follows
\begin{eqnarray}\label{out2}
\phi^{out}_E(r<r_H-\Delta(E))=A_E\phi^{in}_E \exp\left[
2i\int^{r_H-\Delta(E)}_{r_H+\Delta(E)}k'(r)dr\right] \exp\left[ 2i
\int^r_{r_H-\Delta(E)} {Er\over (r-r_H)}dr\right].
\end{eqnarray}
Using eq.(\ref{wave}), the first integration in RHS of above
equation can be calculated,
\begin{eqnarray}\label{k2}
2i\int^{r_H-\Delta(E)}_{r_H+\Delta(E)} k'(r)dr=\pm
2Er_H\int^1_{-1}{dx\over\sqrt{1-x^2}}\sqrt{x+3\xi(x^2-1)+3\xi^2x(x^2-1)+\xi^3(x^4-1)\over
{x(1+x\xi)(1+{\xi\over \xi x^2+x})}}.
\end{eqnarray}
Since $\xi\ll 1$, we have
\begin{eqnarray}
2i\int^{r_H-\Delta(E)}_{r_H+\Delta(E)} k'(r)dr \simeq\pm (2\pi
r_HE\pm i2\pi r_H E\xi+3\pi r_HE\xi^2+O(\xi^3))\simeq \pm 2\pi r_H
E(1\pm i\xi).
\end{eqnarray}
where the terms of $O(\xi^2)$ have been neglected (due to $\xi\ll
1$), whose effects will be briefly discussed in the end of this
paper. Taking the positive sign, we obtain the absolute value of
ratio of the outgoing wave function's amplitude outside the black
hole to the one inside the black hole as follows
\begin{eqnarray}\label{Ra}
\left|{\phi^{out}_E(r>r_H+\Delta(E))\over\phi^{out}_E(r<r_H-\Delta(E))}\right|=
e^{-2\pi r_H E}
\end{eqnarray}
According to the Sannan's discussions\cite{Sannan} and using
eq.(\ref{Ra}), the relative scattering probability from the event
horizon reads
\begin{eqnarray}
P_E=e^{-4\pi r_H E}.
\end{eqnarray}
This means that the $\phi_E$-particle mean number $\langle {\cal
N}_E \rangle $ in the radiation is as follows\cite{Sannan}

\begin{eqnarray}
\langle {\cal N}_E \rangle ={\left|\Gamma_E\right|^2\over e^{4\pi
r_H E}-1}\equiv {\left|\Gamma_E\right|^2\over e^{E/T_H}-1}
\end{eqnarray}
where $|\Gamma_E|^2$ is  the frequency-dependent transmission
coefficient for the outgoing wave to reach future infinity.
Consequently, in the formalism of this present paper we obtain
\begin{equation}\label{T}
\beta^{-1}=T_H={1\over 4\pi r_H},
\end{equation}
which is same as the Hawking temperature.

With the above, let's return to the deriving of the black hole
entropy. Substituting eq.(\ref{T}) into Eq.(\ref{S}), we get the
entropy of the black hole as follows
\begin{eqnarray}\label{bh}
 S={\zeta(3)N\over 16\pi^4}{A\over G}
 \end{eqnarray}
where $A=4\pi r_H^2$ is the  horizon area. So we reproduce the
correct relation of $S \propto A$, which implies that we are on
the right track.  We shouldn't expect that the precise
Bekenstein-Hawking entropy can be obtained via the brick-wall
model even with our semi-classical consideration without fine
tuning certain parameters. Nevertheless, our semi-classical
consideration does provide explanations to the boundary condition
imposed for the field considered and to the brick wall thickness
parameter $h$, therefore, still quite remarkable.

If we set the number of the quantum field multiplet $N={4\pi^4
/\zeta(3)}\approx 324$, then Eq. (\ref{bh}) does give the right
entropy.  The statistical average value of $\Delta(E)$ can be
obtained as
 \begin{eqnarray}\label{D}
 \overline{\Delta(E)}=l_p^2{{\int^\infty_0 {Edg(E)\over e^{\beta
 E}-1}}\over {\int^\infty_0 {dg(E)\over e^{\beta
 E}-1}}}\approx {3\zeta(3)l^2_p\over\pi^3 r_H}
 \end{eqnarray}

The brick wall thickness given in Eq.(1) can be related to the
above $\overline{\Delta(E)}$ via
\begin{equation}\label{hh}
h=\eta \overline{\Delta(E)},\;\;\;\;\eta={N' \pi^2\over
1080\zeta(3) }.
\end{equation}
This indicates that the thickness hypothesis in the brick wall
model reflects the average effect of the noncommutative field
theory in the wall. Since the non-commutativity arises from the
semi-classical consideration, therefore we provide a direct link
of the thickness hypothesis in the model to the underlying quantum
effect as expected.

Finally, we argue that the effects of $O(\xi^2)$ in eq.(\ref{k2})
raise the effective temperature of the hole as it radiates.
Namely, the $\xi$-dependency in the eq.(\ref{k2}) should be
thought as its thermal statistical average
$\overline{\xi}$-dependency. Like eq.(\ref{D}), $\overline{\xi^2}
=l_p^4\overline{E^2}/r_H^2=1/40\times (l_p/r_H)^4$, then the
effective temperature for the hole is
\begin{equation}
T_{\rm eff}=T_H(1-{3 l_p^4 \over 80 r_H^4}).
\end{equation}
where the $T_H$ is Hawking temperature and the second term in the
right-hand-side represents a correction to the temperature due to
the space-time non-commutative property near the event horizon
eq.(\ref{e4}). Obviously, this correction to the $T_H$ is tiny as
$r_H>>l_p$, and hence it can be ignored indeed. The corrections of
$O(\xi^N)$ with $N>2$ can be analyzed likewise, and they are also
ignorable as $r_H>>l_p$.

\begin{center}
{\bf ACKNOWLEDGMENTS}
\end{center}
 The authors wish to
thank Jiliang Jing, Jian-Xin Lu and Shuang-Qing Wu for their
stimulating discussions. We would like also to thank the referee
for his important suggestion on deriving the black hole
temperature in this paper. This work is partially supported by NSF
of China 90103002 and the Grant of the Chinese Academy of
Sciences.

\end{document}